\documentclass[12pt,]{article}

\usepackage{amsmath}
\usepackage{amsfonts} \usepackage{amssymb}

\numberwithin{equation}{section}

\begin{document}      
\title{Time-like hypersurfaces of prescribed mean \\
extrinsic curvature.
}
\author {Helmut Friedrich\\
Max-Planck-Institut f\"ur Gravitationsphysik\\
Am M\"uhlenberg 1\\ 14476 Golm, Germany }
\maketitle

{\footnotesize

 \begin{abstract}
 
 \noindent
The results on the initial boundary value problem for Einstein's vacuum field equation obtained in \cite{friedrich:nagy} rely on an unusual gauge. One of the defining gauge source functions  represents the mean extrinsic curvature of the time-like leaves of a foliation that includes the boundary and covers a neighbourhoood of it. The others steer the development of a frame field and coordinates on the leaves. In general their combined action is needed to control in the context of the reduced field equations the evolution of the leaves.
In this article are derived the hyperbolic equations implicit in that gauge. It is shown that the latter are independent of the Einstein equations and  well defined on arbitrary space-times.
The analysis simplifies if boundary conditions
with constant mean extrinsic curvature  are stipulated. It simplifies  further  if the boundary is required to be totally geodesic.

\end{abstract}




\section{Introduction}

In this article we consider a question that arises in 
the context of the initial boundary value problem for Einstein's vacuum field equation formulated in \cite{friedrich:nagy}. 
The setting is that of a smooth Lorentz metric $g$ on a manifold $M$ with boundary  $S \cup T$ where $S$ and $T$ are smooth hypersurfaces of $M$ which are space- and time-like respectively and intersect in the space-like surface $\Sigma = S \cap T$ that represents the common boundary of $S$ and of $T$. The manifold $T$ is assumed to be diffeomorphic to 
$\mathbb{R}^+_0 \times \Sigma$ with $\{0\} \times \Sigma$ identified with $\Sigma$. 
The assumptions on $S$ are usually chosen according to the desired application.   
For definiteness we assume it to be compact with boundary 
$\Sigma$, though this is not needed for the following arguments. The set $M \setminus S$ is supposed to be in the future of $S$ and on one side of $T$ so that all past directed non-extendible time-like curves in $M$ acquire an endpoint on $S \cup T$. As in \cite{friedrich:ibvp-unique} a smooth space-time $(M, g)$ with these properties is referred to as an {\it $ST$-space-time}. The initial boundary value problem asks for the existence and uniqueness of solutions 
to  Einstein's vacuum field equations $R_{\mu \nu} = 0$
that induce suitably prescribed  Cauchy data  on $S$  and boundary data  on $T$. Such a solution will be called an {\it $ST$-vacuum solution}.

The analysis of the standard Cauchy problem for Einstein's equations is typically based on coordinates $x^{\nu}$
that obey a {\it wave gauge} characterized by  
{\it gauge source functions} $F^{\mu} = F^{\mu}(x^{\nu})$ \cite{friedrich:1hyp red}.
In this gauge, often referred to as `harmonic'  
if $F^{\mu} = 0$ and (somewhat absurdly) as `generalized
harmonic' if $F^{\mu} \neq 0$, 
Einstein's equations take the form of a system of wave equations of second order, the {\it reduced equations}. 
If $g$ is a solution to this system for Cauchy data that satisfy the constraints and the gauge conditions on a space-like initial hypersurface, the reduced equations imply with the Bianchi identity a {\it subsidiary system} which allows one to conclude that the coordinates satisfy the semi-linear system of wave equations  
\begin{equation}
\label{gauge-equ}
\Box_g x^{\mu} = F^{\mu}(x^{\nu}). 
\end{equation}
This ensures that $g$
is in fact a solution to the Einstein equations. 

As pointed out in \cite{friedrich:1hyp red}, the concept of a gauge source function gives access to a huge class of useful gauge conditions.  In principle any coordinate system (ignoring situations of very weak differentiability) can be  used to get hyperbolic reduced equations. If a solution admits coordinates that exist globally, these can be characterized in terms of gauge source functions. Since they combine by (\ref{gauge-equ}) information on the coordinates as well as the metric, the corresponding functions $F^{\mu}$ are usually not known a priori. 
A successful use of the concept thus requires a clever choice of the functions $F^{\mu}(x^{\nu})$ or suitable generalizations thereof. Interesting examples of such applications can be found
 in \cite{ringstroem:2008} in a purely analytical context 
 and in
\cite{lindblom+al:2006}, \cite{pretorius:2005a}, \cite{pretorius:2005b}  in numerical contexts.

The analysis of the initial boundary value problem in \cite{friedrich:nagy} rests on an unusual gauge and  is characterized by  unusual gauge source functions $f$ and $F^A$, $A = 1, 2$. Here $f$ 
corresponds to a function of the connection coefficients that
controls in a neighbourhood of the boundary $T = T_0$ the evolution of a family of time-like hypersurfaces $T_c$, $0  \le c < \epsilon$, which define a smooth space-time foliation that extends into the interior of the solution space-time. 
The two functions $F^A$, $A = 1, 2$, correspond to connection coefficients that
control the evolution of a time-like vector field tangential to the hypersurfaces $T_c$, that serves to define coordinates on these hypersurfaces.

With these particular fields singled out as the gauge source functions  there can be extracted from the Einstein equations, in the representation used in
\cite{friedrich:nagy}, a hyperbolic system of reduced equations that allows one to formulate with suitably prescribed data a well-posed initial boundary value problem. The choice of this system is motivated by the fact that 
Einstein's equations
do not supply evolution equations for $f$ and $F^A$. Its final  justification follows from the existence of a hyperbolic subsidiary system which implies that the reduced system preserves the constraints and gauge conditions and that the latter do what they have been chosen for. 

This leads to a well-posedness result,  local in time, for the initial boundary value problem for Einstein's field equations and one could leave it at that. 
In this article we would like to show, however, that there are explicit equations, analogous to (\ref{gauge-equ}), that reveal the relation between the gauge source functions $f$ and $F^A$, the gauged structures, and the metric. Such equations will show that the desired 
gauge can be established under general assumptions and without any particular equation imposed on $g$. They give, in particular,  additional confirmation  that the gauge conditions do not impose restrictions on the solutions. Moreover, the knowledge of these equations could help to find gauge source functions that extend the life time of the gauge and 
give control on the long term behaviour or  
other desired features of the solutions.

On the solution space-time the restriction of $f$ to $T_c$ represents
 the mean extrinsic curvature induced on this hypersurface and thus encodes an implicitly  evolution law for $T_c$. This is in general not independent of the functions $F^A$.  The mean extrinsic curvature $\chi$ of the boundary $T$ constitutes in the setting of \cite{friedrich:nagy} a boundary datum which can be prescribed freely. Thus $f$ should be smooth and coincide with $\chi$ on $T$ but can be chosen rather arbitrary
 elsewhere. In the special case where  the boundary data are given so that $\chi = \chi_* =const.$ on $T$, one can choose $f =  \chi_*$ on the foliation. As pointed out in \cite{friedrich:nagy}, this leads to a considerable simplification. If 
 $\chi$ is point-dependent, however, $f$ must be point-dependent
and  the part of the gauge controlled by $F^A$ comes into play. This problem is much more involved and has not been analyzed so far. 
 
 The situation is  reminiscent of the problem of constructing standard Cauchy data for Einstein's field equations with point-dependent mean extrinsic curvature. For a long time it was customary to construct Cauchy data with constant mean extrinsic curvature (CMC). The main reason was that this led to technical simplifications but it had even been assumed occasionally  that asymptotically flat solutions always admit such slices \cite{brill:1982}. The analysis of the Einstein evolution equations, however,  does not require  such an assumption on the initial data.
In 1982 Dieter Brill showed
that there exist asymptotically flat solutions to Einstein's field equations which do not admit a Cauchy hypersurface with vanishing mean extrinsic curvature  \cite{brill:1982} 
 and  in 1988 Robert Bartnik showed that there exist cosmological space-times which do not admit CMC Cauchy slices  \cite{bartnik:1988b}. 
Neglecting data with point-dependent mean extrinsic curvature may thus exclude large classes of important space-times.
Only recently have been obtained results on such data of some generality \cite{holst:nagy:tsogtgerel:2008}, \cite{maxwell:2009}.

Similarly, we shall miss out on large classes of space-times developing from initial and boundary data if the mean extrinsic curvature on the
boundary is required to be constant,
a point dependence of $\chi$ could induce the boundary $T$
to shrink or bulge or oscillate. 
The nature and origin of the difficulties arising in the case 
of a point-dependent mean extrinsic curvature on the time-like
boundary are, however, quite different from those arising in the construction of standard Cauchy data on space-like slices.

To explain the way the well posed initial boundary value problem
 is set up in \cite{friedrich:nagy}, some  considerations 
 of \cite{friedrich:nagy}
will be recalled in section 2. 
 In section 3 are discussed various aspects of the special case of constant mean extrinsic curvature $\chi = \chi_*$. This will in particular shed additional light on the result of
 Grigorios Fournodavlos and Jacques Smulevici \cite{fournodavlos-smulevici:2020}, who studied the case of boundaries that are totally geodesic.  
  In the section 4 we finally consider the case of a point-dependent datum $\chi$. It is shown in which sense the functions $f$ and $F^A$
are related to an  implicit quasi-linear, symmetric hyperbolic system that fixes the gauge.

In the context of the initial boundary value problem long term evolution projects are beset with all the difficulties known from the standard Cauchy problem. There are issues, however, that are specific to this problem. 
If a solution $(M, g)$ to Einstein's equation is given, a manifold of the form $T = \mathbb{R}^+_0 \times \Sigma$ can be smoothly embedded as a time-like hypersurface but it can also just  be immersed so that the pull-back of $g$ still defines a smooth Lorentzian metric on $T$ but the image shows self-intersections. The ambient space-time then still induces on $T$ smooth vacuum data of the type considered below.
 We can thus imagine  situations where 
$T$ is smoothly embedded close to its initial boundary $\Sigma$ but then, as the space-time evolves, opposite sides of $T$ start to move towards each other and threaten to touch and intersect.
The point is that the  data on $T$ alone may not allow one to decide whether they are induced by an embedding or an immersion  and even if they are induced by an embedding they may represent situations 
where discrete points of $T$ are mapped to points in $M$ 
lying quite close to each other.
If one is ambitious enough to analyze such situations the gauge problem may need reconsiderations. In the following we shall not be ambitious and only consider the initial boundary value problem local in time.

\section{The setting.}

In the following we consider $4$-dimensional $ST$-space-times $(M, g)$ with boundary $S \cup T$ and edge $\Sigma = S \cap T$
as described in the introduction.
Our goal is to construct solutions to Einstein's vacuum 
field equation $R_{\mu\nu}[g] = 0$ that arise from suitably prescribed initial data on $S$ and boundary data on $T$.
On the domain of dependence of the initial hypersurface $S$ they are uniquely  determined by the standard Cauchy problem for Einstein's vacuum field equation with Cauchy data on $S$. 
Seeking to construct local in time solutions to initial boundary value problems, a first step is to  control the solutions in a neighbourhood of the edge $\Sigma$ that simultaneously represents the boundary of 
$S$ and of $T$. The choice of data on the boundary $T$ depends very much on the chosen representation of  the field equations.
We follow here the discussion of  \cite{friedrich:nagy}, which employs a frame formalism and uses the vacuum Bianchi equations for the conformal Weyl tensor.

\subsection{Formalism and gauge conditions.}

On any smooth ST-space-time $(M, g)$ can be chosen as follows 
an {\it $ST$-adapted gauge}, consisting of a smooth coordinate system $x^{\mu}$, $ \mu = 0, \,. \,. \, , 3$, and smooth local  orthonormal  frames $e_k$, $k = 0, \,. \,. \, , 3$ near $T$.
 
\vspace{.1cm}

The function $x^3$ satisfies $x^3 = 0$ on $T$, $x^3 > 0$ elsewhere,  so that the sets $T_c = \{x^3 = c = const.\}$ 
with $0 \le c < c_*$ and $T_0 = T$ are  time-like hypersurfaces diffeomorphic to $T$ that smoothly foliate some  neighbourhood $W$ of $T$ in $M$ with $dx^3 \neq 0$ on $W$.
 
\vspace{.1cm}

The unit vector field  $e_0$ on $W$ is time-like, future directed, tangential to the hypersurfaces $T_c$, and 
orthogonal to the 2-surfaces $S_c = S \cap T_c$
with $S_0 = \Sigma$. 
The space-like vector field $e_3$ on $W$ represents the inward directed unit normals to the $T_c$.

\vspace{.1cm}

The local vector fields $e_A$, $A = 1, 2$, define together with  $e_0$ and $e_3$ a local orthonormal frame on some open subset of $S \cap W$, so that
\[
g(e_i, e_j) = g_{ij} = \eta_{ij} = \mbox{diag}(1, - 1, -1, -1).
\] 
The $e_A$ are then tangential to the surfaces $S_c$.
Denoting by $D$ the Levi-Civita connection of the Lorentz-3-metric induced by $g$ on $T_c$ we require the fields $e_A$
to be $D$-Fermi transported in the direction of $e_0$ so that
\[
g(e_0, e_0)\,D_{e_0}\,e_A + g(e_A,D_{e_0}\,e_0)\,e_0 
- g(e_A,e_0)\,D_{e_0}\,e_0 = 0.
\]
The fields $e_A$ are then everywhere tangential to $T_c$.

\vspace{.1cm}

The function $x^0$ defines a natural parameter on the integral curves of $e_0$ 
so that
\begin{equation}
\label{x0-nat-par}
<e_0, dx^{0}>\, = 1 \quad \mbox{on} \quad W,
 \quad x^0 = 0 \quad \mbox{on} \quad S \cap W.
\end{equation}
Consider the vector field  $X = (q^{\#}(dx^3, dx^3))^{-1}\,\mbox{grad}_qx^3$ tangential to $S$ and orthogonal to the $S_c$, where 
$q$ denotes the metric induced on $S$  by $g$.
 An integral curve $\gamma(\sigma)$ of $X$
with $\gamma(0) \in \Sigma$ satisfies
\[
\frac{d}{d\sigma}(x^3(\gamma(\sigma)))
= \,< dx^3, \frac{d}{d\sigma}\gamma(\sigma)>\,
= \,< dx^3, (q^{\#}(dx^3, dx^3))^{-1}\,\mbox{grad}_qx^3>\, = 1,
\]
so that
$\sigma = x^3(\gamma(\sigma))$, whence $\gamma(c) \in S_c$.
The flow of $X$ thus maps $\Sigma$ diffeomorphically onto the $S_c$
and we have a parametrization  
$W = \{(x^0, p, x^3)|\,\,x^0 \ge 0, \,\,p \in \Sigma, \,\,0 \le x^3 < c_*\}$.
 Choose local coordinates $x^{\alpha}$, $\alpha = 1, 2$, on  $\Sigma$,
assume them to be dragged into the interior of $S$ with the flow of $X$  so that 
\begin{equation}
\label{x-alpha-in-S}
q^{\#}(dx^{\alpha}, dx^3) = 0, \quad \alpha = 1, 2, 
\end{equation}
and then
dragged along with the flow of $e_0$ so that  
\begin{equation}
\label{x-alpha-in-W}
<e_0, dx^{\alpha}>\, = 0  \quad \mbox{on} \quad W, 
 \quad \alpha = 1, 2.
\end{equation}
The $x^{\alpha}$, $\alpha = 1, 2$, define local coordinates on the $S_c$ and the $x^{\alpha}$, $\alpha = 0, 1, 2$,
define local coordinates on $T_c$ for $0 \le c < c_*$. 
For the frame coefficients satisfying $e_k = e^{\mu}\,_k\,\partial_{\mu}$ holds 
\begin{equation}
\label{frame-coeff}
e^{\mu}\,_0 = \delta^{\mu}\,_0,
\quad \,\,\,
e^3\,_A = 0, 
\quad \,\,\,
e^3\,_3 > 0 \quad  
\mbox{on \,\,$T_c$ \,\,and} \quad e^0\,_A = 0, \quad A = 1, 2,
\quad \mbox{on \,\,$S_c$}.
\end{equation}

We will have to consider three types of projections.
Since our frame is well adapted to the geometrical situation,
corresponding projection formalisms can be avoided by distinguishing three
groups of indices. They are given, with the values they take, by
\[
a, c, d, e, f = 0, 1, 2; \quad \quad i,j,k,l,m,n = 0, 1, 2, 3;
\quad \quad  A, B, C, D = 1, 2. 
\]
For each group the summation rule is assumed. If $\nabla$ denotes the connection defined by $g$, the connection coefficients
$\Gamma_j\,^i\,_k$ in the frame $e_k$ satisfy
$\nabla_je_k \equiv \nabla_{e_j}e_k = \Gamma_j\,^i\,_k\,e_i$ and 
$\Gamma_{jlk} = - \Gamma_{jkl}$
with $\Gamma_{jlk} = g_{li}\,\Gamma_j\,^i\,_k$.
The second fundamental form induced on $T_c$
in the frame $e_a$ and the mean extrinsic curvature of the hypersurfaces $T_c$ are given by
\[
\chi_{ab} \equiv g(\nabla_{e_a}\,e_3, e_b) 
= - g(e_3, \nabla_{e_a}\,e_b) 
= \Gamma_a\,^3\,_b = \Gamma_{(a}\,^3\,_{b)},
\]
\begin{equation}
\label{mec}
\chi \equiv g^{ab}\,\chi_{ab} = g^{jk}\,\Gamma_j\,^3\,_k = 
\nabla_{\mu}\,e^{\mu}\,_3,
\end{equation}
respectively. Because
\begin{equation}
\label{ftagauge}
D_a\,e_c \equiv D_{e_a}\,e_c = \Gamma_a\,^b\,_c\,e_b,
\end{equation}
the $ \Gamma_a\,^b\,_c$ define the inner connection $D$ on
$T_c$.
The Fermi condition implies 
\begin{equation}
\label{Fgauge}
\Gamma_0\,^A\,_B = 0,
\end{equation}
\begin{equation}
\label{frame-equ}
D_{e_0}\,e_0 = F^A\,e_A,\quad
D_{e_0}\,e_A = - F_A\,e_0 \quad \mbox{with} \quad 
F^A = \Gamma_0\,^A\,_0,  \quad F_A = \eta_{AB}\,F^A.
\end{equation}

\vspace{.1cm}

The freedom of choosing the function $x^3$ and the time-like vector field $e_0$ we started with finds new expression  in this formalism. With $e_a$ as given above
the functions $F^A = F^A(x ^{\mu})$ follow from these formulas. We can, however, also think of the functions $F^A$ as being at our free disposal. If we solve 
equation (\ref{frame-equ}) on $T_c$ with arbitrarily prescribed functions $F^A = F^A(x ^{\alpha}, c)$ and initial data satisfying
$g(e_a, e_b) = \eta_{ab}$ on $S_c$ and $e_0$ orthogonal to $S_c$, the solution will satisfy the relation $g(e_a, e_b) = \eta_{ab}$ on $T_c$. Since the field equations 
(\ref{torf}), (\ref{Gcurv}), (\ref{frvacbian})
do not provide propagation equations for
the $F^A$ this suggests to consider  these functions  in the reduced field equations as smooth gauge source function that can be freely prescribed.
With the special choice
\begin{equation}
\label{e0-geodesic}
F^A = 0, 
\end{equation}
the field $e_0$ will be $D$-geodesic and the fields $e_A$ parallel propagated. More general choices of $F^A$ may allow one to avoid the development of caustics.

The freedom of choosing the leaves $T_c$ of the foliation defined by the function $x^3$ is encoded in the mean extrinsic curvature $\chi = \chi(x^{\alpha}, c)$ induced on $T_c$. It will be discussed below in detail how $T_c$ is determined by $\chi = \chi(x^{\alpha}, c)$ and the initial data on $S \cap W$. Since equations 
(\ref{torf}), (\ref{Gcurv}), (\ref{frvacbian})
do not provide a propagation equation for $\chi$ this function will be considered as a  gauge source function that will be represented by a smooth function $f = f(x^{\mu})$  in the reduced field equations. There is, however, a slight difference with the $F^A$. The function $\chi(x^{\alpha}, 0)$ will be used as a free boundary datum on $T$ that determines the form of $T$. While the function $f$ can be freely extended into the interior of the solution space-time, it must thus be given so that $f(x^{\alpha}, 0) = \chi(x^{\alpha}, 0)$ on $T$.

\vspace{.1cm}

Due to the compactness of $\Sigma$, whence of $S_c = S \cap T_c$,
the coordinates  $x^{\alpha}$, $\alpha = 1, 2$ and, in general, also the frame vector fields 
$e_A$ are only defined locally on $S_c$. 
 If they have to be redefined on some overlap region $U$
in $W \cap S$ it is important to note that the corresponding transformation is explicitly controlled along the integral curves of $e_0$ by the propagation laws imposed above and the transformation on $U$.

If the frame is subject to a transformation 
$e_A \rightarrow e_{A'} = s^A\,_{A'}\,e_A$ in some overlap patch $U \subset S \cap W$ with some point dependent transformation $s^A\,_{A'} \in SO(2)$ on $S_c$, we must require
\[
F^A \rightarrow  F^{A'} = s^{A'}\,_B\,F^B \,\,\, \mbox{whence}  \,\,\, 
F_A \rightarrow F_{A'} = s^A\,_{A'}\,F_A
 \,\,\, \mbox{with}  \,\,\,  s^A\,_{A'}\,s^{A'}\,_B = \delta^A\,_B, 
\]
to preserve the first of equations (\ref{frame-equ}) and the second equation then requires
\[
D_{e_0}s^A\,_{A'} = 0,
\]
which allows us to control the corresponding transformations
\[
\eta_{AB} \rightarrow  \eta_{A'B'} = \eta_{AB}\,s^A\,_{A'}\,s^B\,_{B'}, \quad
\beta_{AB} \rightarrow  \beta_{A'B'} = \beta_{AB}\,s^A\,_{A'}\,s^B\,_{B'},
\]
along the integral curves of $e_0$.


In the coordinates $x^{\alpha}$, $\alpha = 1, 2$ described above 
$f$ is is assumed to be smooth with
 local representation $f = f(x^0, x^{\alpha}, x^3)$
on $W$.
 If the coordinates are subject on a subset of $\Sigma$ to a transformation
$x^{\alpha} \rightarrow x^{\beta'} =  x^{\beta'}(x^{\alpha})$
with inverse $x^{\alpha} = x^{\alpha}(x^{\beta'})$, this transformation transfers by the rules
(\ref{x-alpha-in-S}) and (\ref{x-alpha-in-W})  into $W$, which allows us to control explicitly the new coordinate representation  
$f' = f'(x^0, x^{\alpha'}, x^3) = f(x^0, x^{\beta}( x^{\alpha'}), x^3)$.
In a similar way transform the coordinate representations of the functions $F^A$ on $W$ and of the functions $\chi$, $\eta_{AB}$,
$\beta_{AB}$, $\alpha$, and $\beta$ on $T$ discussed below.

\subsection{The field equations.}

The basic unknowns in the representation of the field equations used in 
\cite{friedrich:nagy} are 
\begin{equation}
\label{complete-unknown}
e^{\mu}\,_k,\,\,\,\,\,\Gamma_k\,^i\,_j,\,\,\,\,\,\,C^{i}\,_{jkl},
\end{equation}
where $C^{i}\,_{jkl}$ is a tensor field with the algebraic properties of a  conformal Weyl tensor which for a solution of the equations will
in fact assume that meaning. The field equations are given by the {\it torsion free condition}
\begin{equation}
\label{torf}
[e_i,e_j] = (\Gamma_i\,^k\,_j - \Gamma_j\,^k\,_i)\,e_k,  
\end{equation}
where the square bracket denotes the commutator of the vector fields,
the {\it curvature relation}
\begin{equation}
\label{Gcurv}
e_k(\Gamma_l\,^i\,_j) - e_l(\Gamma_k\,^i\,_j) + 
2\,\Gamma_{[k}\,^i\,_{|m|} \Gamma_{l]}\,^m\,_j - 
2\,\Gamma_m\,^i\,_j \,\Gamma_{[k}\,^m\,_{l]} = 
C^i\,_{jkl},
 \end{equation}
where the left hand side gives the curvature of the connection $\nabla$ in terms of the connection coefficients and the frame, and the {\it vacuum Bianchi identity}
\begin{equation}
\label{frvacbian}
\nabla_{i}\,C^{i}\,_{jkl} = 0.
\end{equation}

\subsubsection{Splittings of the conformal Weyl tensor.}

 In the following we shall need two different decompositions of the conformal Weyl tensor.
 The decomposition used in \cite{friedrich:nagy} is defined by the time-like frame vector field $n = e_0$.
{\it The $n$-electric} and {\it the $n$-magnetic} part   
of the conformal Weyl tensor are given by 
\[
E^n_{ik} = p_i\,^m\,p_k\,^n\,C_{mjnl}\,n^j\,n^l, \quad \quad 
B^n_{ik} = 
p_i\,^m\,p_k\,^n\, \frac{1}{2}\,C_{mjpq}\,\epsilon^{pq}\,_{nl}\,n^j\,n^l,
\]
respectively, where 
$\epsilon_{ijkl} =  \epsilon_{[ijkl]}$ with $\epsilon_{0123} = 1$
and $p_{ij} = g_{ij} - n_i\,n_j$. 
These tensors are symmetric, trace free, and {\it spatial} in the sense that 
$n^i\,E^n_{ik} = 0$ and $n^i\,B^n_{ik} = 0$. It holds
\[
C_{ijkl} = 2 \left(q_{j[k}\,E^n_{l]i} - q_{i[k}\,E^n_{l]j} 
- n_{[k}\,B^n_{l]m}\,\epsilon^{m}\,_{ij} 
- n_{[i}\,B^n_{j]m}\,\epsilon^{m}\,_{kl} \right),
\] 
with
$q_{ij} = g_{ij} - 2\,n_i\,n_j$ and 
$\epsilon_{jkl} = n^i\,\epsilon_{ijkl} = \epsilon_{0jkl}$, and also
\[
C_{mnpq}\,n^m\,p^n\,_{j}\,p^p\,_{k}\,p^q\,_{l} 
= C_{0npq}\,p^n\,_{j}\,p^p\,_{k}\,p^q\,_{l} 
= - B^n_{jm}\,\epsilon^m\,_{kl},
\]
\[
C_{mnpq}\,p^m\,_{i}\,p^n\,_{j}\,p^p\,_{k}\,p^q\,_{l} 
= 2 \left(p_{j[k}\,E^n_{l]i} - p_{i[k}\,E^n_{l]j}\right).
\] 
The second decomposition is defined by the space-like frame vector field $N = e_3$.
{\it The $N$-electric} and {\it the $N$-magnetic} part   
of the conformal Weyl tensor are given by 
\[
E^N_{ik} = k_i\,^m\,k_k\,^n\,C_{mjnl}\,N^j\,N^l, \quad \quad 
B^N_{ik} = 
k_i\,^m\,k_k\,^n\, \frac{1}{2}\,C_{mjpq}\,\epsilon^{pq}\,_{nl}\,N^j\,N^l.
\]
These tensors are symmetric, trace free, satisfy 
$N^i\,E^N_{ik} = 0$,  $N^i\,B^N_{ik} = 0$, and  
\[
C_{ijkl} = 2 \left(- l_{j[k}\,E^N_{l]i} + l_{i[k}\,E^N_{l]j} 
+ N_{[k}\,B^N_{l]m}\,\bar{\epsilon}^{m}\,_{ij} 
+ N_{[i}\,B^N_{j]m}\,\bar{\epsilon}^{m}\,_{kl} \right),
\] 
where
$k_{ij} = g_{ij} + N_i\,N_j$,   
$l_{ij} = g_{ij} + 2\,N_i\,N_j$,
$\bar{\epsilon}_{ijk} = \epsilon_{ijkl}\,N^l
 = \epsilon_{ijk3}$.
 It holds
\[
C_{mnpq}\,N^m\,k^n\,_{j}\,k^p\,_{k}\,k^q\,_{l} 
= C_{3npq}\,k^n\,_{j}\,k^p\,_{k}\,k^q\,_{l} 
= - B^N_{jm}\,\epsilon^m\,_{kl},
\]
i.e.
\[
C^3\,_{abc} = B^N\,_{ad}\,\epsilon^d\,_{bc3}
\quad   \mbox{or} \quad 
B^N\,_{ad} = \frac{1}{2}\,C_{3abc}\,\epsilon^{bc}\,_{3d},
\]
and
\[
C_{mnpq}\,k^m\,_{i}\,k^n\,_{j}\,k^p\,_{k}\,k^q\,_{l} 
= 2 \left(- k_{j[k}\,E^N_{l]i} + k_{i[k}\,E^N_{l]j}\right).
\] 
The different parts are related by
\[
B^N_{00} = C_{0312} = B^n_{33}, \quad \quad
B^N_{01} = C_{0302} = E^n_{23}, \quad \quad 
B^N_{02} = C_{0310} = - E^n_{13}, \quad \quad 
\]
\[
B^N_{10} = C_{1312} = E^n_{32}, \quad \quad 
B^N_{11} = C_{1302} = - B^n_{22}, \quad \quad 
B^N_{12} = C_{1310} = B^n_{12}, \quad \quad 
\]
\[
B^N_{20} = C_{2312} = - E^n_{13}, \quad \quad 
B^N_{21} = C_{2302} = B^n_{12}, \quad \quad 
B^N_{22} = C_{3201} = - B^n_{11}, \quad \quad 
\]

\[
E^N_{00} = C_{0303} = E^n_{33}, \quad \quad 
E^N_{01} = C_{0313} = - B^n_{32}, \quad \quad 
E^N_{02} = C_{0323} = B^n_{31}, \quad \quad 
\]
\[
E^N_{10} = C_{1303} = - B^n_{32}, \quad \quad 
E^N_{11} = C_{1313} = - E^n_{22}, \quad \quad 
E^N_{12} = C_{1323} = E^n_{12}, \quad \quad 
\]
\[
E^N_{20} = C_{2303} = B^n_{31}\, \quad \quad 
E^N_{21} = C_{2313} = E^n_{21}, \quad \quad 
E^N_{22} = C_{2323} = - E^n_{11}. \quad \quad 
\]

\subsubsection{The Gauss - Codazzi equations}

The tensor 
\[
k_{ij} = g_{ij} + N_i\,N_j = \eta_{ab}\,\delta^a\,_i\,\delta^b\,_j,
\]
represents the metric induced on the hypersurfaces $T_c$. We shall need the well known equations which relate the curvature tensor $R^i\,_{jkl}[g]$ of the metric $g$ to fields living on the hypersurfaces $T_c$.
Gauss' equation, which reads in our formalism 
\begin{equation}
\label{gauss}
\bar{R}^a\,_{bcd}[k] = R^a\,_{bcd}[g] 
- \chi_c\,^a\, \chi_{d b} 
+ \chi_d\,^a\, \chi_{cb},
\end{equation}
relates it to the curvature tensor $ \bar{R}^a\,_{bcd}[k] $ of $k$ and the second fundamental form.
Codazzi's equation
\begin{equation}
\label{codazzi}
D_c\chi_{db} - D_d\chi_{cb} = R^3\,_{bcd}[g],
\end{equation}
relates it to $T_c -$intrinsic derivatives of the second fundamental form.
With the well known decomposition of 
 of the curvature tensor on $3$-dimensional spaces and the relations on the Weyl tensor above, Gauss' equation can be written on vacuum solutions
\[
 k_{b[d}\,R_{c]a}[k]+ k_{a[c}\,R_{d]b}[k]
 +  \frac{1}{3}\,R[k]\,k_{a[d}\,k_{c]b}
\]
\[
= 
2 \left(k_{b[d}\,E^N_{c]a} + k_{a[c}\,E^N_{d]b}\right)
 - \chi_{ca}\, \chi_{d b} 
 +  \chi_{da}\, \chi_{cb}.
 \]
where $R_{ab}[k]$ and $R[k]$ denote the Ricci tensor and the Ricci scalar of $k$. A contraction gives
\begin{equation}
\label{contr-vac-gauss}
\frac{1}{2}\,\left(R_{db}[k]  +  \frac{R[k]}{3}\,k_{db}\right) = E^N_{db} 
- \chi\,\chi_{bd} + \chi_d\,^c\,\chi_{bc}.
\end{equation}
Codazzi's equation takes on a vacuum solution the form
\begin{equation}
\label{vac-codazzi}
D_c\chi_{db} - D_d\chi_{cb} = C^3\,_{bcd} = B^N\,_{be}\,\epsilon^e\,_{cd3}.
\end{equation}

\subsubsection{Reduced equations and boundary data.}

In the following a few remarks  will be made on the reduced equations, the initial and boundary data, the corner conditions, and the subsidiary equations. For details (which are sometimes slightly rewritten here) the reader is referred to  \cite{friedrich:nagy}. 

\vspace{.1cm}

With the gauge conditions and the gauge source functions $F^A = F^A(x^{\mu})$ and $f = f(x^{\mu})$ equations (\ref{torf}), (\ref{Gcurv}), (\ref{frvacbian}) imply a symmetric hyperbolic system of {\it reduced equation}
for the unknowns
\[
e^{\beta}\,_a, \quad e^{\mu}\,_3, \quad 
\Gamma_A\,^B\,_0, \quad \Gamma_A\,^B\,_C, \quad 
\Gamma_3\,^A\,_B, \quad \Gamma_3\,^A\,_0, \quad
\Gamma_3\,^3\,_A, \quad\Gamma_3\,^3\,_0, 
\]
\[
\chi_{01}, \quad  \chi_{02}, \quad \chi_{11}, \quad \chi_{12}, \quad 
\chi_{22}, \quad B^n_{ab}, \quad E^n_{ab}.
\]
The field $\chi_{00}$, which only occurs in non-differentiated form in these equations, is taken care of  by writing $\chi_{00} = \chi_{11} +\chi_{22} + f$.

\vspace{.2cm}

The {\it initial data} on $S$ are given by {\it standard Cauchy data}, i.e. a solution to the vacuum constraints on space-like hypersurfaces, which extend smoothly to the boundary $\Sigma$ of $S$.

\vspace{.2cm}

To describe the {\it boundary conditions} we consider
the trace free parts of the orthogonal projections of the 
$n$-electric and $n$-magnetic parts of the conformal Weyl tensor on $T$ into the planes orthogonal to $e_3$ and $e_0$.
The corresponding symmetric trace free tensors
$\eta_{AB}$ and $\beta_{AB}$, which by the relations given above can also be expressed in terms of the  $N$-electric and $N$-magnetic parts  of the conformal Weyl tensor, are given
 by  
\begin{equation}
\label{betaAB-comp}
\beta_{11}  = - \beta_{22} = - \frac{1}{2}\,(C_{3201} + C_{3102}) 
= \frac{1}{2}\,( B^n_{11} - B^n_{22}) = \frac{1}{2}\,(B^N_{11} - B^N_{22}),
\end{equation}
\[
\beta_{12} = \beta_{21} = C_{3101} = B^n_{12} = B^N_{12},
\]
and 
\begin{equation}
\label{etaAB-comp}
\eta_{11} = - \eta_{22} =  \frac{1}{2}\,(C_{1010} - C_{2020})  = \frac{1}{2}\,(E^n_{11} - E^n_{22}) 
 = \frac{1}{2}\,(E^N_{11} - E^N_{22}), 
\end{equation}
\[
\eta_{12} = \eta_{21} = E_{12} = C_{1020}  = E^n_{12}  = E^N_{12}. 
\]
In terms of the pseudo-orthonormal frame $l$, $k$, $m$
satisfying (with $e_3$ inward pointing) 
\[
\sqrt{2}\,l = e_0 + e_3,\,\,\,\,\,\, 
\sqrt{2}\,k = e_0 - e_3,\,\,\,\,\,\, 
\sqrt{2}\,m = e_1 - i\,e_2,
\]
and the Newman-Penrose notation for the curvature tensor, the relevant components
of the conformal Weyl tensor are given
 by
\[
\Psi_0 = C_{\mu\nu\sigma\pi}\,l^{\mu}\,m^{\nu}\,l^{\sigma}\,m^{\pi}
= \eta_{11} + \beta_{12} + i\,(\beta_{11} - \eta_{12}),
\]
\[
\Psi_4 = C_{\mu\nu\sigma\pi}\,\bar{m}^{\mu}\,k^{\nu}\,\bar{m}^{\sigma}\,k^{\pi}
= \eta_{11} - \beta_{12} + i\,(\beta_{11} + \eta_{12}).
\]
The boundary conditions of \cite{friedrich:nagy} then take the form
\begin{equation}
\label{cbdrycond}
f = \chi, \quad
- \Psi_4 + \alpha\,\Psi_0 + \beta\,\bar{\Psi}_0 = q
\quad \mbox{on} \quad T,
\end{equation}
where $\alpha$ and  $\beta$ are complex-valued functions 
on $T$ that satisfy
\begin{equation}
\label{explHcond} 
 |\alpha| + |\beta| \le 1,
\end{equation}
and $q$, the main datum besides 
$\chi$ on $T$, is a smooth complex-valued function on $T$ that can be prescribed, consistent with the conditions discussed below, freely. 
Condition (\ref{explHcond}), which looks simpler than the corresponding condition given in \cite{friedrich:nagy}, is obtained from the latter by diagonalizing the matrix $B$ used  there to express the restrictions on  $\alpha$ and  $\beta$.

\vspace{.2cm}

As special examples, which will be of interest below, 
we note that the admissible choice $\alpha = 0$ and $\beta = 1$ results in the boundary  condition
\begin{equation}
\label{beta-bdry-cond}
q = 2\,(\beta_{12} - i\,\beta_{11}) = 2\,B^n_{12} - i\,(B^n_{11} - B^n_{22}) 
 = 2\,B^N_{12} - i\,(B^N_{11} - B^N_{22})
 \quad \mbox{on} \quad T.
\end{equation}
It only involves magnetic parts. The choice
$\alpha = 0$, $\beta = - 1$ gives the boundary condition
\begin{equation}
\label{eta-bdry-cond}
q = - 2\,(\eta_{11} + i\,\eta_{12}) = E^n_{22} - E^n_{11} - 2\,i\,E^n_{12}
= E^N_{22} - E^N_{11} - 2\,i\,E^N_{12}
 \quad \mbox{on} \quad T,
\end{equation}
which only involves electric parts
of the conformal Weyl tensor.

\vspace{.2cm}

To determine a smooth solution, the initial and the boundary data must satisfy a consistency condition, the so-called {\it corner condition}  at the edge $\Sigma$. Because the reduced equations with given $f$ and $F^A$ are symmetric hyperbolic, the Cauchy data on $S$ determine in our gauge a unique formal expansion type solution on $S = \{x^0 = 0 \}$. 
For given functions $\alpha$ and $\beta$ this expansion determines, in particular,  a unique formal expansion of the fields on the left hand sides of (\ref{cbdrycond}) at $\Sigma$ . 
The corner conditions require the data $\chi$ and $q$ on the right hand sides to be prescribed consistent with these expansions  at $\Sigma$.

For given Cauchy data it is always possible to find boundary data that satisfy this condition, which leaves  the right hand sides of  (\ref{cbdrycond}) essentially arbitrary away from $\Sigma$. If the boundary data are supposed to satisfy certain conditions, possibly suggested by some intended application, it requires an extra effort (and may not be possible) 
to construct Cauchy data, i.e. solutions to the constraint equations on $S$, that meet these requirements.

The main result of  \cite{friedrich:nagy} says (in the notation introduced above): 

\noindent
{\it Let be given smooth Cauchy data for Einstein's vacuum field equations on the compact $3$-manifold $S$ with boundary $\Sigma$ 
and smooth boundary data 
\begin{equation}
\label{gen-boundary-data}
\chi, \,\,\, 
q \quad \mbox{and functions}  \quad 
\alpha, \,\,\beta
\quad 
\mbox{on} \quad T  = \mathbb{R}^+ \times \Sigma,
\end{equation} 
satisfying  (\ref{explHcond}). 
Choose on $M = \mathbb{R}^+ \times S$ smooth gauge source 
functions $f$ with $f = \chi$ on $T$ and
$F^A $ that satisfy together with the initial and boundary data the corner conditions at $\Sigma$ defined by the reduced field equations. 
Then there exists for some $x^0_* > 0$ a unique smooth solution $g$ to Einstein's vacuum equations on the manifold 
$M' = [0,  x^0_* [ \,\times \, S \subset \mathbb{R}^+ \times S$ 
so that $S \equiv \{0 \} \times S$ is space-like, 
$T' = [0,  x^0_* [ \,\times \, \Sigma \subset T$ is time-like, $g$ 
 induces the given Cauchy  data on $S$,  
$\chi$
acquires on $T'$ the meaning of the mean intrinsic curvature induced on $T'$ by $g$  and $q$ coincides with the  function of the conformal Weyl tensor of $g$ on the left hand side of  the second equation of (\ref{cbdrycond}).}

\vspace{.2cm}

The first step to arrive at this result consists in setting up local initial boundary value problems for the reduced field equations and showing the existence of local solutions near given points of $\Sigma$. In a next step  the local 
solutions are patched  together to obtain a solution covering a neighbourhood of $\Sigma$. Because, as pointed out above, the transformations between local solutions can be explicitly controlled, there arises no problem. The solution near  $\Sigma$ is then  patched together with the solution to the Cauchy problem for the reduced equations that is determined by the data on $S$.

 This establishes the existence of a unique solution to the reduced equations on a manifold of the form $M' = [0,  x^0_* [ \,\times \, S$.
In a final step a hyperbolic {\it subsidary system} is derived that
supplies an argument that the solution to the reduced equations is in fact a solution to the Einstein equations.

\vspace{.2cm}

Suppose $(M, g)$ is a $ST$-vacuum solution. After choosing a gauge as above one can read off the data induced on $S$ and $T$ and the gauge source functions $f$ and $F^A$ near $T$ (the corner conditions will, of course, be satisfied). Our result then shows that local in time the given $ST$-vacuum solution will be reconstructed uniquely by our method. Thus local in time all  $ST$-vacuum solutions are covered by the existence result  above.

\section{Time-like hypersurfaces and mean extrinsic curvature.}

In this section we discuss the basic equation associated with the 
mean extrinsic curvature on time-like hypersurfaces. We consider then the case of constant mean extrinsic curvature and derive the equations which are in this case implicit in the formulation of the  initial boundary value problem of \cite{friedrich:nagy}. It follows a discussion of totally geodesic boundaries.

\subsection{The basic equation.}

Let $({\cal M}, g)$ be a $4$-dimensional space-time, ${\cal S}$ a space-like hypersurface which we assume for convenience to be a Cauchy hypersurface of ${\cal M}$, and ${\cal T}$ a time-like hypersurface which intersects ${\cal S}$ so that it cuts out from it a compact $3$-dimensional manifold $S$ with compact space-like boundary $\Sigma = {\cal S} \cap {\cal T}$. 
The sets $S$ and $\Sigma$ may be thought of as identical with the ones labeled 
by the same symbols in the previous section.
The set ${\cal T}$ does represent a space-time boundary but serves as a subsidiary hypersurface to establish a certain equation. It is fairly arbitrary, only when we arrive at condition (\ref{k00>0-cond}) we will need to restrict it, together with the coordinates $z^{\mu}$ introduced below, further.
The metric $g$ is not required to satisfy any field equation.
In the following the coordinate indices $\alpha, \beta, \gamma, \delta$
take
values  $0, 1, 2$ while the coordinate indices $\kappa, \lambda, \mu, \nu, \pi,  \rho$ 
take values  $0, 1, 2, 3$.

\vspace{.1cm}

Let $z^0$, $z^3$   with  $dz^0 \neq 0$, $dz^3 \neq 0$ denote smooth functions
defined on a neighbourhood of $\Sigma$ so that ${\cal S} = \{z^0 = 0\}$,  ${\cal T} = \{z^3 = 0\}$, and $z^3 > 0$ on $S$ on that neighbourhood. 
Set $S_c = \{z^0 = 0,\,\,z^3 = c = const.\}$ for $0 \le c < c_*$ so that 
$S_0 = \Sigma$. By our assumptions we have  
$g^{00} = g^{\#}(dx^0,\,dx^0) > 0$ on ${\cal S}$,   
$g^{33} = g^{\#}(dx^3,\,dx^3) < 0 $ on ${\cal T}$, and thus 
$g^{00} > 0$ and $g^{33}  < 0$ on the $S_c$ for suitably chosen $c_* > 0$. In the following this will always be assumed.

The functions $z^0$, $z^3$ are complemented by functions  $z^1$, $z^2$
so that the $z^{\mu}$ define local coordinates,
$z^1$, $z^2$, $z^3$ define local coordinates on $S$,
$z^1$, $z^2$ define local coordinates on the $S_c$, and 
 $z^0$, $z^1$, $z^2$ define local coordinates on ${\cal T}$.
The vector field $\partial_{z^0}$ is future directed.

\vspace{.1cm}

To construct a time-like hypersurface $T_c$ with 
$T_c \cap {\cal S} = S_c$ whose mean extrinsic curvature coincides with a given function $\chi$ we assume it to be  
essentially given as the graph of a function 
$\phi = \phi(z^{\alpha})$ over ${\cal T}$ so that
\[
T_c = \{\Phi(z^{\mu}) = c\} \quad \mbox{where} 
 \quad \Phi(z^{\mu}) = z^3 - \phi(z^{\alpha})
 \quad \mbox{with} \quad  
\]
\begin{equation}
\label{phi-in-value}
\phi|_{z^0 = 0} = 0 \quad \mbox{whence} \quad 
\phi_{, \alpha}|_{z^0 = 0} = \delta^0\,_{\alpha}\,\phi_{,0} 
\quad  \mbox{on} \quad S_c.
\end{equation}
For $T_c$ to be time-like we need
\begin{equation}
\label{phi,0-in-value}
\nabla_{\nu}\,\Phi\,\nabla^{\nu}\,\Phi
= g^{33} - 2\,g^{3 0}\,\phi_{,0} 
+ g^{0 0}\,\phi_{,0}\,\phi_{,0} < 0
\quad \mbox{on} \quad S_c.
\end{equation}
This inequality  is satisfied if
\begin{equation}
\label{phi,0-range}
\frac{g^{03}}{g^{00}} - \sqrt{- \frac{g^{33}}{g^{00}} + 
\left(\frac{g^{03}}{g^{00}}\right)^2} < \phi_{,0} <
\frac{g^{03}}{g^{00}} + \sqrt{- \frac{g^{33}}{g^{00}} + 
\left(\frac{g^{03}}{g^{00}}\right)^2}.
\end{equation}
Since $g^{33} < 0$ and $g^{00} > 0$ on $S_c$
 the roots are real and the unit normal to $T_c$ will be well defined near $S_c$ and given by the restriction to  $T_c$ of the vector field
\begin{equation}
\label{N-normal}
N^{\mu} =\nu\,\nabla^{\mu}\,\Phi = \nu\,
(g^{\mu 3} - g^{\mu \alpha}\phi_{,\alpha}),
\end{equation}
with
\[ 
\nu = - (- \nabla_{\nu}\,\Phi\,\nabla^{\nu}\,\Phi)^{- \frac{1}{2}}
= - (-  g^{33} + 2\,g^{3 \alpha}\,\phi_{, \alpha} 
- g^{ \alpha  \beta}\,\phi_{, \alpha}\,\phi_{, \beta})^{- \frac{1}{2}}. 
\]
The gradient fields in brackets on the right hand side of 
(\ref{N-normal}) are orthogonal to $\{x^3 = c\}$ and ${\cal S}$ respectively and thus to the tangent spaces $T_pS_c$  of $S_c$.
With $\phi_{,0}$ varying in the range (\ref{phi,0-range}) 
the vector $N^{\mu}$ exhausts all space-like directions in the 2-plane orthogonal to the tangent space $T_pS_c$
of $S_c$. That plane contains the time-like vector
\begin{equation}
\label{T-ortho-N-and-Sc}
T^{\mu} = 
(g^{03} - \phi_{,0}\,g^{00})\,g^{\mu 3}
- (g^{33} - \phi_{,0}\,g^{03}) g^{\mu 0},
\end{equation}
which is orthogonal to $N$ and $T_pS_c$ and unique up to a factor. 
It follows with  (\ref{phi,0-in-value})
\begin{equation}
\label{T0-positive}
g^{33} - \phi_{,0}\,g^{03} < \frac{1}{2}\,(
g^{33} - g^{0 0}\,\phi^2_{,0}) 
< 0
\quad \mbox{on $\,S_c\,$ near $\,\Sigma$}.
\end{equation}
The metric induced on $T_c$ is given by
\begin{equation}
\label{k-mu-nu}
k_{\mu \nu} = g_{\mu \nu} + N_{\mu}\,N_{\nu},
\end{equation}
the induced second fundamental form is given by 
\begin{equation}
\label{chi-mu-nu}
\chi_{\mu \nu} = 
k_{\mu}\,^{\kappa}\,k_{\nu}\,^{\lambda}\,\nabla_{\kappa}\,N_{\lambda}
= \nu\,k_{\mu}\,^{\kappa}\,k_{\nu}\,^{\lambda}\,
\nabla_{\kappa}\,\nabla_{\lambda}\,\Phi,
\end{equation}
\[
= 
- \nu\,k_{\mu}\,^{\kappa}\,k_{\nu}\,^{\lambda}\,(
\phi_{,\alpha \beta}\,\delta^{\beta}\,_{\kappa}\,\delta^{\alpha}\,_{\lambda}
+ \Gamma_{\kappa}\,^{\rho}\,_{\lambda}
(\delta^3\,_{\rho} - \phi_{,\alpha}\,\delta^{\alpha}\,_{\rho}))
\]
and the mean extrinsic curvature by 
\[
\chi = k^{\mu\nu}\,\chi_{\mu\nu} = 
k^{\mu\nu}\,\nabla_{\mu}\,N_{\nu} =
\nabla_{\mu}\,N^{\mu} = 
\nu\,k^{\mu\nu}\,\nabla_{\mu}\,\nabla_{\nu}\,\Phi.
\]
The equation which controls the evolution of $\phi$ and thus of $T_c$ thus reads
\begin{equation}
\label{phi-equ}
- \nu\,k^{\alpha \beta} \,\partial_{\alpha}\,\partial_{\beta}\,\phi
- \nu\,k^{\mu\nu}\,(
\Gamma_{\mu}\,^3\,_{\nu}
- \Gamma_{\mu}\,^{\alpha}\,_{\nu}\,\phi_{,\alpha})
= \chi,
\end{equation}
where here and in the following the background fields $g_{\mu\nu}$
$\Gamma_{\mu}\,^{\rho}\,_{\nu}$ as well as
$\nu$, $k^{\mu \nu}$ etc.
have to be taken at the points $(z^{\alpha}, z^3) 
= (z^{\alpha}, \phi(z^{\alpha}) + c)$. 

In the following we will consider the $z^{\alpha}$ also as coordinates on 
$T_c$. A vector field on $T_c$ will then be of the form 
$X = X^{\alpha}\,\partial_{z^{\alpha}}$. Considered as a vector field in 
$M$ is must be written $X' = X^{\alpha}\,\partial_{z^{\alpha}} + 
\phi_{,\alpha}\,X^{\alpha}\,\partial_{z^{3}}$. It holds then
$g(X', Y') = \bar{k}(X, Y)$ with $ \bar{k}$ 
the pull back $\bar{k}$ of $k_{\mu\nu}\,dz^{\mu}\,dz^{\nu}$ to 
the hypersurface $T_c = \{(z^{\mu}) = (z^{\alpha}, \phi(z^{\alpha}) + c)\}$. It coincides with the pull back
 of $g_{\mu\nu}\,dz^{\mu}\,dz^{\nu}$ to $T_c$ and is given by 
\begin{equation}
\label{bar(k)-down-alpha-beta}
\bar{k} = \bar{k}_{\alpha \beta}\,dz^{\alpha}\,dz^{\beta} =
(g_{\alpha \beta} + 2\,g_{3 (\alpha}\,\phi_{, \beta)} +
g_{33}\,\phi_{,\alpha}\,\phi_{,\beta})\,dz^{\alpha}\,dz^{\beta}.
\end{equation}
As long as $T_c$ is time-like this metric is Lorentzian.

The principal part of equations (\ref{phi-equ}) is governed 
by the symmetric tensor
\begin{equation}
\label{bar(k)-up-alpha-beta}
k^{\alpha \beta} = \bar{k}^{\alpha \beta} \equiv g^{\alpha \beta}
 + \nu^2\,(g^{\alpha 3} - g^{\alpha \gamma}\,\phi_{,\gamma})
\,(g^{\beta 3} - g^{\beta \delta}\,\phi_{,\delta})
\quad \mbox{on} \quad T_c, 
\end{equation}
with 
\[
\nu^2 = (- g^{33} + 2\,g^{3 \alpha}\,\phi_{,\alpha} - 
g^{\alpha \beta}\phi_{,\alpha}\,\phi_{, \beta})^{- 1}.
\] 
It satisfies 
\[
\bar{k}_{\alpha \beta}\, \bar{k}^{\beta \gamma} = \delta_{\alpha}\,^{\beta}.
\]
Suppose $\chi$ is  is a given function of four variables.
We write then $\chi = \chi(z^{\alpha}, \phi(z^{\alpha}) + c)$
on the right hand side of (\ref{phi-equ}). 
{\it Equation (\ref{phi-equ}) defines then a quasi-linear wave equation for $\phi$.
With a given, sufficiently small constant $c$,
a  given right hand side $\chi$ as above, and
initial data $\phi$ and $\phi_{,0}$ on $S_c$ that satisfy
 (\ref{phi-in-value}) and (\ref{phi,0-in-value}) 
 it  determines a unique solution $\phi$ near $S_c$
 for which the hypersurface 
$T_c = \{z^3 = \phi(z^{\alpha}) + c\}$ is  time-like.
Because of the smooth dependence of the solutions on $c$, the initial data and $\chi$, the hypersurfaces $T_c$
represent the leaves of a smooth local foliation of ${\cal M}$ near 
$\Sigma$ that defines a smooth function 
$x^3$ with $x^3 = c$ on $T_c$.}

\subsubsection{Boundaries of constant mean extrinsic curvature.}

The coordinate dependence of the function $\chi$ has been specified above in a somewhat cursory way. In the context of the initial boundary value problem  it requires in general further considerations. 
There is, however, a case, pointed out already  in  \cite{friedrich:nagy}, where things simplify considerably. If the data  (\ref{cbdrycond}) are given with
\begin{equation}
\label{chi-const-boundary-data}
\chi = \chi_* = \mbox{constant on $T$},
\end{equation} 
we can set $f =  \chi_*$ near $T$.
Equation (\ref{phi-equ}) simplifies and
the construction of the hypersurfaces $T_c$ and the coordinate $x^3$  
completely decouples from setting up the 
frame vectors $e_a$ and coordinates $x^{\alpha}$
and thus from the choice of the $F^A$.
The reduced equations will ensure that their solution satisfies $\chi = \chi_*$ near $T$.

Simple though not uninteresting examples of such situations are given by  the hypersurfaces 
$T = \{r = const. > 2\,m\}$
of the Schwarzschild solution in standard Schwarzschild coordinates.
Their mean extrinsic curvature is given by 
\begin{equation}
\label{S1-chi}
\chi =
-  \sqrt{1 - \frac{2\,m}{r}}\,\left(\frac{2\,r - 3\,m}{r\,(r - 2\,m)}\right).
\end{equation}
This still leaves  the freedom to prescribe as a datum the function $q$ in (\ref{cbdrycond}) on $T$. Its deviation from the Schwarzschild values may be thought of as representing ingoing or outgoing gravitational radiation.
The situation changes if we consider the  Kerr solution in Boyer-Lindquist coordinates. The mean extrinsic curvature of the 
hypersurface $T = \{r = const. > 2\,m\}$ is then given by
 \[
\chi =
- \frac{1}{2}\,\sqrt{\frac{r^2 - 2\,m\,r + a^2}{r^2 + a^2\,\cos^2\theta}}
\left\{\frac{2\,(r - m)}{r^2 - 2\,m\,r + a^2} 
+ \frac{2\,r}{r^2 + a^2\,\cos^2\theta}
\right\}.
\]

\subsubsection{Totally geodesic boundaries}

An even more restricted case is considered by Grigorios Fournodavlos and Jacques Smulevici \cite{fournodavlos-smulevici:2020}, who study the existence of solutions with boundaries that are totally geodesic,  emphasizing that the problem of geometric uniqueness discussed in \cite{friedrich:ibvp-unique} is absent here.
We consider this situation here again because the present formalism sheds additional light onto the special nature of these issues in this case.

In general we are free to prescribe three functions on the boundary. On the face of it, the condition $\chi_{ab} = 0$, needed to make the boundary totally geodesics,  thus looks much too strong. Precisely because all components of $\chi_{ab}$ are required to vanish, 
it turns out, however,   
that the problem can be reduced to the prescription of three functions.
Suppose $(M, g)$ is a $ST$ vacuum solution with totally geodesic time-like boundary $T$ so that $\chi_{ab} = 0$ on $T$. 
To set up a gauge as described in section 2, 
we can use equation (\ref{phi-equ}) with $\chi = 0$ on the right hand side to construct the function $x^3$, which has been handpicked before, so that $f = 0$ near $T$. For convenience we can then require 
(\ref{e0-geodesic}) near $T$.

The Codazzi equation 
(\ref{vac-codazzi}) implies with our assumption that
\begin{equation}
\label{B-N=0}
B^N\,_{ab} = 0 \quad \mbox{on} \quad T,
\end{equation} 
and thus the, by 
(\ref{beta-bdry-cond}) admissible, boundary conditions
\begin{equation}
\label{tot-geod-data}
\chi = 0, \quad
\beta_{12} = B^N_{12} = 0,
 \quad
\beta_{11} 
 = \frac{1}{2}\,(B^N_{11} - B^N_{22}) = 0
\quad \mbox{on} \quad T.
\end{equation}
Together with the Cauchy data induced by $g$ on $S$  these boundary conditions determine the solution uniquely near $S$.
Because the solution is smooth, the corner conditions are satisfied, which guarantees that
\begin{equation}
\label{chiab-on-Sigma}
\chi_{ab} = 0 \quad \mbox{on} \quad \Sigma.
\end{equation}

But then the data (\ref{tot-geod-data}) and (\ref{chiab-on-Sigma}) lead us back to where we started from.
The reduced equations derived in \cite{friedrich:nagy} 
comprise under general assumptions  the subsystem
\[
D_0\,\chi_{01} - D_1\,\chi_{11} - D_2\,\chi_{12} = D_1(f), 
\]
\[
D_0\, \chi_{02} - D_1\,\chi_{12} - D_2\,\chi_{22} = D_2(f), 
\]
\[
\quad \quad\quad \,\,\,\, D_0\,\chi_{11} - D_1\,\chi_{01} = - \beta_{12},  
\]
\[
2\,D_0\,\chi_{12} - D_1\,\chi_{02} - D_2\,\chi_{01}
= 2\,\beta_{11},\quad 
\]
\[
\quad \quad\quad\,\,  D_0\,\chi_{22} - D_2\,\chi_{02} = \beta_{12}, 
\]
where $f = \chi_{00} - \chi_{11} - \chi_{22}$, 
$\beta_{11}  = \frac{1}{2}\,( B^n_{11} - B^n_{22})
 = \frac{1}{2}\,(B^N_{11} - B^N_{22})$
and  $\,\,\beta_{12} = B^n_{12} = B^N_{12}$. 
Because the vector fields $e_a$ are tangential to $T$  
this subsystem defines a system intrinsic to  the boundary $T$.
It is symmetric hyperbolic. With the boundary condition
(\ref{tot-geod-data}) the right hand sides of the equations vanish on $T$. Observing the consistency condition (\ref{chiab-on-Sigma}), we can conclude, without explicit knowledge of the frame and the connection coefficients on $T$, that $\chi_{ab} = 0$ on $T$.

\vspace{.2cm}

The argument shows that every $ST$-vacuum solution 
with totally geodesic boundary $T$ can be obtained locally in time
by solving the initial boundary value problem with boundary conditions that satisfy (\ref{tot-geod-data}) and Cauchy data that imply (\ref{chiab-on-Sigma}).  
The main problem of characterizing all such solutions then reduces to the construction of Cauchy data for Einstein's vacuum field equations on $3$-manifolds $S$ with boundary $\Sigma$ for which the reduced equations determine a formal expansion with $\chi_{ab} = 0$,
$B^N\,_{ab} = 0$ on $\Sigma$. 

The particular choice of the gauge source functions made here is convenient but nowhere enters the argument. 
In fact, because of (\ref{B-N=0}) conditions  (\ref{tot-geod-data}) will be satisfied in any frame. The obstructions to geometric uniqueness pointed out in \cite{friedrich:ibvp-unique} simply do not occur in this particular case.
Anyway, the discussion of
\cite{friedrich:ibvp-unique} does not ask so much for specific cases in which geometric uniqueness may hold but refers to the general problem with the complete freedom of prescribing three functions as boundary data.

In the case of anti-de Sitter type solutions that admit a smooth conformal boundary ${\cal J}$ at space-like and null infinity, the boundary it totally geodesic in a suitable conformal gauge as consequence of the field equations. 
Nevertheless, there is the freedom, in a sense similar to the second of conditions (\ref{cbdrycond}),
 to freely prescribe  on ${\cal J}$ two functions derived from the conformal Weyl tensor. Moreover, with a condition similar to 
(\ref{beta-bdry-cond}) the boundary conditions can be stated in a completely geometric way \cite{friedrich:AdS}.

Since the Codazzi equation so much simplifies the above argument  one may wonder whether something similar could be done by imposing conditions on the metric induced on $T$ and 
using Gauss' equation (\ref{contr-vac-gauss}). Because of the occurrence of the second fundamental form in that relation
there appears to be no obvious way.

\section{Prescribed mean extrinsic curvature.}

The case of point dependent mean intrinsic curvature is complicated because the function  $\chi$ (or the gauge source function $f$) is not given
 in terms of some arbitrary coordinates like the  $z^{\alpha}$ considered above. A relation like:  $\chi = \chi(x^{\alpha'}, c)$ on $T_c$ or  $f = f(x^{\alpha'}, c)$ on $T_c$  in section 2  is saying that we must think of $\chi$ or $f$ as being given in the specific coordinates $x^{\alpha'}$  on $T_c$
 that are  obtained by solving equations (\ref{x0-nat-par}), (\ref{x-alpha-in-W}), (\ref{frame-equ})  on $T_c$. 
 {\it We write $x^{\alpha'}$ here to distinguish the index coming with $x$ from the completely unrelated index of $z^{\alpha}$.} While the function $f$ can be chosen freely away from the boundary $T = T_0$, the free data must be specified on $T$ in some distinguished coordinate system. This
 coupling between the gauge on $T$ and the way boundary data are prescribed, which is a specific feature of the initial boundary value problem, cannot be avoided unless the boundary conditions are completely stated in terms geometric structures (see the discussion in \cite{friedrich:ibvp-unique}). The functions 
 $F^A = F^A(x^{\alpha'})$ will have to play a role when we specify $\chi$ or $f$.

Equations (\ref{x0-nat-par}), (\ref{x-alpha-in-W}),
(\ref{frame-equ}), i.e.
\begin{equation}
\label{e-x-evol}
D_{e_0}\,x^{\alpha'} = \delta^{\alpha'}\,_0,
\quad \quad
D_{e_0}\,e_0 = F^A\,e_A,
\quad \quad 
D_{e_0}\,e_A = - F_A\,e_0,
\end{equation}
are defined in terms of structures supposed to be  induced by the ambient space-time on the hypersurface $T_c$ which we want to determine by solving equation (\ref{phi-equ}).
The right hand side of the latter must thus be arranged so as to correspond in the coordinates $x^{\alpha'}$ to  the datum $\chi = \chi(x^{\alpha'}, c)$  (or  $f = f(x^{\alpha'}, c)$) assumed as known. 

This suggests to consider  $F^A(x^{\alpha'}, c)$, $f(x^{\alpha'}, c)$, 
$ \chi(x^{\alpha'},c )$ as given functions and use
(\ref{e-x-evol}) with $F^A = F^A(x^{\alpha'}(z^{\beta}, c), c)$
as a system of equations for 
$x^{\alpha'} = x^{\alpha'}(z^{\beta}, c)$ and the frame vectors $e_a$.
The latter are thought to be given in the form 
$e^{\alpha}\,_a\,\partial_{z^{\alpha}}$ with
$e^{\alpha}\,_a = e^{\alpha}\,_a(z^{\beta})$ 
and can be expressed in terms of the coordinates $x^{\alpha'}$ once the coordinate transformation is available
on $T_c$. 
The system (\ref{e-x-evol}) must be coupled to (\ref{phi-equ}) 
with $\chi = \chi(x^{\alpha'}(z^{\beta}, c), c)$
or $f = f(x^{\alpha'}(z^{\beta}, c), c)$ on the right hand side to determine $\phi$ and thus $T_c$. The dependence of the various functions on $c$ will often be suppressed in the following because we mostly work with a fixed $c$,.

This recipe does not work immediately. Two of equations (\ref{e-x-evol}) 
involve the Levi-Civita connection $D$ of the metric 
$\bar{k}_{\alpha \beta}$  on $T_c$ by (\ref{bar(k)-down-alpha-beta}), and thus the Christoffel symbols
\begin{equation}
\label{k-bar-Chris-symb}
\bar{\zeta}\,_{\alpha}\,^{\beta}\,_{\gamma} =
\frac{1}{2}\,\bar{k}^{ \beta \,\delta}\,
(\bar{k}_{\delta \gamma, \alpha} + \bar{k}_{\alpha \delta,  \gamma} 
- \bar{k}_{\alpha \gamma, \delta})
\quad \mbox{on} \quad T_c.
\end{equation}
They depend on 
the functions $g_{\mu\nu}(z^{\alpha}, \phi(z^{\alpha}) + c)$, their derivatives 
$g_{\mu\nu, \beta}(z^{\alpha}, \phi(z^{\alpha}) + c)
+ g_{\mu\nu, 3}(z^{\alpha}, \phi(z^{\alpha}) + c)\,\phi_{,\beta}$,
on the $\phi_{, \alpha}$, and in particular on the second derivatives 
$\phi_{,\alpha \beta}$. 
To control these functions we write equation
(\ref{phi-equ}) in the form
\begin{equation}
\label{short-phi-equ}
 - \nu\,k^{\alpha \beta} \,\partial_{\alpha}\,\partial_{\beta}\,\phi
- F
=   \chi(x^{\alpha'}(z^{\beta})),
\end{equation}
with
\[
F =
\nu\,k^{\mu\nu}\,(
\Gamma_{\mu}\,^3\,_{\nu}
- \Gamma_{\mu}\,^{\alpha}\,_{\nu}\,\phi_{,\alpha})
= F(\phi, \phi_{,\alpha}, \mbox{background}) ,
\]
where  the background function depend on  $(z^{\alpha},\phi(z^{\alpha}) + c)$. To obtain an equation that supplies the $\phi_{,\alpha \beta}$,
we apply $\partial_{z^{\gamma}}$ to the equation above and obtain
\begin{equation}
\label{phi,alpha-equ}
- \nu\,k^{\alpha \beta} \,\partial_{\alpha}\,\partial_{\beta}\,
(\partial_{\gamma}\,\phi)
- G
= \partial_{x^{\alpha'}}\chi\,\,\partial_{z^{\gamma}}x^{\alpha'}
\end{equation}
with some
\[
G = G(\phi, \phi_{,\alpha}, \phi_{,\alpha \beta}, \mbox{background}). 
\]
where the background functions now involve  
$(z^{\alpha}, \phi(z^{\alpha}), \phi_{,\beta}(z^{\alpha}))$. 
For given smooth functions $x^{\alpha'}(z^{\beta})$ 
the two equations above define a system of wave equations
 for $\phi$ and $\phi_{,\alpha}$.  
In addition to $x^{\alpha'}(z^{\beta})$ 
the right hand side of (\ref{phi,alpha-equ}) requires, however,
 also control of  
\begin{equation}
\label{dx-dz}
\partial_{z^{\gamma}}\,x^{\alpha'}
= D_{\gamma}x^{\alpha'}.
\end{equation}
 The solvability of the coupled system 
(\ref{e-x-evol}), (\ref{short-phi-equ}), (\ref{phi,alpha-equ})
thus depends on the possibility
to implement equations for this field.
The first of equations (\ref{e-x-evol}) implies
\[
0 = D_{\gamma}\,(e^{\alpha}\,_0\,D_{\alpha}\,x^{\beta'})
= e^{\alpha}\,_0\,D_{\gamma}\,D_{\alpha}\,x^{\beta'}
+ D_{\gamma}\,e^{\alpha}\,_0\,D_{\alpha}\,x^{\beta'}
\]
\[
= D_{e_0}\,(D_{\gamma}\,x^{\beta'})
+ D_{\gamma}\,e^{\alpha}\,_0\,(D_{\alpha}\,x^{\beta'}).
\]
To obtain equations for  
$D_{\gamma}\,e^{\alpha}\,_a$ we use
\[
D_{e_0}\,(D_{\gamma}\,e^{\beta}\,_a) =
e^{\alpha}\,_0\,D_{\gamma}\,D_{\alpha}\,e^{\beta}\,_a
+ R^{\beta}\,_{\delta \alpha \gamma}[\bar{k}]\,
e^{\alpha}\,_0\,e^{\delta}\,_a
\]
\[
=
D_{\gamma}\,(D_{e_0}\,e^{\beta}\,_a)
- (D_{\gamma}\, e^{\alpha}\,_0)\,(D_{\alpha}\,e^{\beta}\,_a)
+ R^{\beta}\,_{\delta \alpha \gamma}[\bar{k}]\,
e^{\alpha}\,_0\,e^{\delta}\,_a.
\]
For the first term on the right hand side we get from
equations (\ref{e-x-evol})  
\[
D_{\gamma}\,(D_{e_0}\,e^{\beta}\,_0)
= F^A\,_{, x^{\alpha'}}\,(D_{\gamma}\,x^{\alpha'})\,e^{\beta}\,_A
+ F^A\,(D_{\gamma}\,e^{\beta}\,_A),
\]
\[
D_{\gamma}\,(D_{e_0}\,e^{\beta}\,_A)
= - F_A\,_{, x^{\alpha'}}\,(D_{\gamma}\,x^{\alpha'})\,e^{\beta}\,_0
- F_A\,(D_{\gamma}\,e^{\beta}\,_0). 
\]
Thus together
\begin{equation}
\label{De0-gamma-x-a'-equ}
D_{e_0}\,(D_{\gamma}\,x^{\beta'})
= - D_{\gamma}\,e^{\alpha}\,_0\,(D_{\alpha}\,x^{\beta'}),
\quad \quad  \quad  \quad
\end{equation}
\begin{equation}
\label{De0-D-gamma-e-0-equ}
D_{e_0}\,(D_{\gamma}\,e^{\beta}\,_0) 
=
F^A\,_{, x^{\alpha'}}\,(D_{\gamma}\,x^{\alpha'})\,e^{\beta}\,_A
+ F^A\,(D_{\gamma}\,e^{\beta}\,_A)  \quad \quad  \quad  \quad 
 \,
\end{equation}
\[
 \quad \quad \quad 
- (D_{\gamma}\, e^{\alpha}\,_0)\,(D_{\alpha}\,e^{\beta}\,_0)
+ R^{\beta}\,_{\delta \alpha \gamma}[\bar{k}]\,
e^{\alpha}\,_0\,e^{\delta}\,_0,
\]
\begin{equation}
\label{De0-D-gamma-e-A-equ}
D_{e_0}\,(D_{\gamma}\,e^{\beta}\,_A) 
=
- F_A\,_{, x^{\alpha'}}\,(D_{\gamma}\,x^{\alpha'})\,e^{\beta}\,_0
- F_A\,(D_{\gamma}\,e^{\beta}\,_0) \quad \quad  \,\,\, 
\end{equation}
\[
 \quad \quad  \quad \quad  - (D_{\gamma}\, e^{\alpha}\,_0)\,(D_{\alpha}\,e^{\beta}\,_A)
+ R^{\beta}\,_{\delta \alpha \gamma}[\bar{k}]\,
e^{\alpha}\,_0\,e^{\delta}\,_A.
\]
These equations
involve the curvature tensor of $\bar{k}$.
Because the Christoffel symbols depend on $\phi_{, \alpha \beta}$ we can expect  third derivatives of $\phi$ to enter 
the expression for the curvature tensor of $\bar{k}$.
Because the background is known, however, we can use 
Gauss' equation 
\begin{equation}
\label{Gauss-equ}
R^{\beta}\,_{\delta \alpha \gamma}[\bar{k}]\,e^{\alpha}\,_0\,e^{\delta}\,_a =
\end{equation}
\[
\left(
\sum_{\rho, \pi, \mu, \nu = 0}^3
R^{\rho}\,_{\pi \mu \nu}[g]\,k^{\beta}\,_{\rho}
k^{\pi}\,_{\delta}\,k^{\mu}\,_{\alpha}\,k^{\nu}\,_{\gamma}
- \chi^{\beta}\,_{\alpha}\,\chi_{\delta \gamma}
+ \chi^{\beta}\,_{\gamma}\,\chi_{\delta \alpha}
\right)e^{\alpha}\,_0\,e^{\delta}\,_a,
\]
where the background field 
$R^{\rho}\,_{\pi \psi \nu}[g]$ is taken at  
$(z^{\alpha}, \phi(z^{\alpha}) + c)$ and (\ref{k-mu-nu}),
(\ref{chi-mu-nu}) are used. 
The terms in large brackets only contain $\phi_{, \alpha}$ and 
$\phi_{, \alpha \beta}$.

With (\ref{Gauss-equ}) taken into account, 
the system
(\ref{e-x-evol}), (\ref{short-phi-equ}), (\ref{phi,alpha-equ}),
(\ref{De0-gamma-x-a'-equ}), (\ref{De0-D-gamma-e-0-equ}),  
(\ref{De0-D-gamma-e-A-equ}) provides the desired closed system for the unknowns
\begin{equation}
\label{unknowns}
x^{\alpha'}, \quad x^{\beta'}\,_{,\alpha}, \quad e^{\beta}\,_a, \quad 
D_{\alpha}\,e^{\beta}\,_a, \quad 
\phi, \quad \phi_{, \alpha}.
\end{equation}

\vspace{.2cm}

\noindent
The initial data on $S_c$ are determined, respectively chosen, as follows.

\vspace{.1cm}

As in (\ref{phi-in-value}) we assume
\[
\phi|_{S_c} = 0 \quad \mbox{so that} \quad 
\phi_{, \alpha} = \delta^0\,_{\alpha}\,\phi_{,0} 
\quad  \mbox{on} \quad S_c.
\]
The function $\phi_{,0}$ must be chosen on  $S_c$ so as to satisfy
(\ref{phi,0-range}). Later on we will be led to consider the further condition (\ref{k00>0-cond}).

The unique future directed unit vector field orthogonal to $N$ and $S_c$ is given by 

\[
e^{\mu}\,_0 = \frac{T^{\mu}}{\sqrt{ (g_{\nu \rho}\,T^{\nu}\,T^{\rho})}}
\quad \mbox{on} \quad S_c,
\]
where $T^{\mu}$ is the vector field (\ref{T-ortho-N-and-Sc}) on $S_c$ which will be tangential to $T_c$. 
Because
$e^3\,_0 = \phi_{,0}\,e^0\,_0 = \phi_{,\alpha}\,e^{\alpha}\,_0$, $e^{\mu}\,_0$  is uniquely determined by 
\[
e^{\alpha}\,_0 = \frac{T^{\alpha}}{\sqrt{ (g_{\nu \rho}\,T^{\nu}\,T^{\rho})}}
= \frac{T^{\alpha}}{\sqrt{ (\bar{k}_{\alpha \beta}\,T^{\alpha}\,T^{\beta})}}
\]
\[
= \nu\,\frac{(g^{03} - \phi_{,0}\,g^{00})\,g^{\alpha 3}
- (g^{33} - \phi_{,0}\,g^{03}) \,g^{\alpha 0}}
{\sqrt{(g^{03})^2 - g^{33}\,g^{00}}}
\quad \mbox{on} \quad S_c.
\]
It satisfies
\[
e^{0}\,_0 
= \nu\,\sqrt{(g^{03})^2 - g^{33}\,g^{00}} > 0.
\]

The fields $e^{\alpha}\,_A$ are chosen tangential to $S_c$ so that
$e^{0}\,_A = 0$, whence 
$\bar{k}_{\alpha \beta}\,e^{\alpha}\,_0\,e^{\beta}\,_A = 0$. 
They are required to satisfy
\[ 
\bar{k}_{\alpha \beta}\,e^{\alpha}\,_A\,e^{\beta}\,_B =
\bar{k}_{C D}\,e^{C}\,_A\,e^{D}\,_B =
g_{C D}\,e^{C}\,_A\,e^{D}\,_B =
 - \delta_{AB}.
\]
The forms dual to $e_a$ are denoted by 
$\sigma^a = \sigma^a\,_{\alpha}\,dz^{\alpha}$ so that
$\sigma^a\,_{\alpha}\,e^{\alpha}\,_b = \delta^a\,_b$. We assume
\[
x^{0'} = 0, \quad \mbox{whence} \quad x^{0'}\,_{,A} = 0 
\quad \mbox{on} \quad S_c.
\]

With the $z^A$ defining local coordinates  on $S_c$, we choose
choose local coordinates $x^{A'} = x^{A'}(z^A)$ on $S_c$,
which give $x^{A'}\,_{,A}$ on $S_c$ with $\det(x^{A'}\,_{,A}) \neq 0$.
Following (\ref{frame-coeff}) we need to require
$\delta^{\alpha'}\,_0 = e^{\alpha'}\,_0 = e^{\alpha}\,_0\,
x^{\alpha'}\,_{, \alpha}$
which implies 
\[
 x^{0'}\,_{,0} =  \frac{1}{e^0\,_0},
  \quad \quad 
 x^{A'}\,_{,0} = -  \frac{1}{e^0\,_0}\,e^A\,_0\,x^{A'}\,_{,A},
\quad \quad e^{0'}\,_A = 0.
\]

It remains to determine
\[
D_{\alpha}\,e^{\beta}\,_b
= \sigma^a\,_{\alpha}\,D_{e_a}\,e^{\beta}\,_b.
\]
By (\ref{frame-equ}) we must set,
 with the right hand sides given by the fields obtained so far,
\[
D_{e_0}\,e^{\beta}\,_0 = F^A\,e^{\beta}\,_A,
\quad \quad 
D_{e_0}\,e^{\beta}\,_A = - F_A\,e^{\beta}\,_0
\quad \mbox{on} \quad S_c.
\]
With the given information we can finally calculate  
\[
D_{e_A}\,e^{\beta}\,_0, 
\quad \quad 
D_{e_A}\,e^{\beta}\,_B 
\quad \mbox{on} \quad S_c,
\]
by using the Christoffel symbols (\ref{k-bar-Chris-symb}).
These involve the functions $\phi_{,\alpha \beta}$ on $S_c$.
From the data given above we get
 \[
\phi_{,AB} = 0 \quad  
\phi_{, 0B}  = (\phi_{, 0})_{,B}
\quad \mbox{on} \quad S_c.
 \]
 To obtain $\phi_{,00}$
we solve (\ref{phi-equ}) on $S_c$ for 
$k^{\alpha \beta}\,\phi_{,\alpha \beta}$ 
and observe that by (\ref{bar(k)-up-alpha-beta}) 
\[
k^{00} = \bar{k}^{00} \equiv g^{00}
 + \nu^2\,(g^{0 3} - g^{00}\,\phi_{,0})^2 > 0 
\quad \mbox{on} \quad S_c. 
\]

\vspace{.2cm}

\noindent
To discuss the solvability of our system we consider its principal part. 

\vspace{.2cm}

The first four of the unknowns (\ref{unknowns}) can be 
combined to an $\mathbb{R}^{j}$-valued unknown $w$ that satisfies an equation of the form
\[
e^{\alpha}\,_0\,w_{, \alpha} = \dots\,,
\]
while the remaining unknowns combine to an $\mathbb{R}^{l}$-valued unknown $u$ that satisfies an equation of the form
\[
\bar{k}^{\alpha \beta}\,u_{,\alpha \beta} = \ldots\,.
\]
That $e_0$ is time-like with respect to $\bar{k}_{\alpha \beta}$
suggests that
the coupled system can be written
as a quasi-linear, symmetric hyperbolic system of first order
\cite{friedrichs:sym-hyp}. 

In fact, 
in terms of the auxiliary unknowns
$v_{\alpha} \equiv u_{, \alpha}$
the 
equation for $u$ implies the system of first order
\[
u_{, 0} = v_{0},
\]
\[
\bar{k}^{00}\,v_{0, 0} 
+ 2\,\bar{k}^{0 A}\,v_{0, A} 
+ \bar{k}^{A B}\,v_{A, B} 
= \ldots
\]
\[
- \bar{k}^{AB}\,v_{B, 0} + \bar{k}^{AB}\,v_{0, B} = 0,
\]
and the equations for $w$ writes
\[
e^{0}\,_0\,w_{,0} 
+ e^{A}\,_0\,w_{,A} 
= \dots\,
\]
We have seen above that $e^{0}\,_0 > 0$ and  $\bar{k}^{00} > 0$ 
on (whence near) $S_c$.
If $\bar{k}^{A B}$ defines a negative definite
symmetric bilinear form 
on (whence near) $S_c$, it follows that the combined 
system is quasi-linear, symmetric hyperbolic. It also implies then for its solutions the integrability condition
$v_{B, 0} = v_{0, B}$ and thus
$(v_{A, B} - v_{B,A})_{,0} = v_{0, AB} - v_{0, BA} = 0$. With suitably given initial data the remaining integrability conditions follow.
It remains to see under which conditions $\bar{k}^{A B}$ is negative definite.

\vspace{.2cm}

{\it Assume
$a, b \in \mathbb{R}$, 
$y, z \in \mathbb{R}^k$, 
$k \ge 2$, and
$A, B$ are 
$k \times k$ matrices so that 
the following matrix equation holds
with $k \times k$ unit matrix $1_k$
\[
\left(
\begin{array}{ccc}
a & ^t y\\
y & A
\end{array}
\right)
\left(
\begin{array}{ccc}
b & ^t z\\
z & B
\end{array}
\right)
=
\left(
\begin{array}{ccc}
1 & ^t0_k\\
0_k & 1_k
\end{array}
\right).
\]
If $b  > 0$ and $A$ is negative definite and
symmetric , i.e. $^tu\,A\,u < 0$ for $u \neq 0$ and
$^tA = A$, then 
$B$ is negative definite and
symmetric if and only if $a > 0$.}

\vspace{.2cm}

In fact, the matrix equation is equivalent to the relations
 \[
 b\,y + A\,z = 0, \quad
 a\,b + \,^ty\,z = 1, \quad 
 y\,^tz + A\,B = 1_k,
\quad a\,^tz + \,^ty\,B = 0.
 \]
The second equation implies $a > 0$ if $y = 0$. Assume that 
$y \neq 0$.
 Being symmetric and negative definite, $A$ has an inverse $A^{-1}$,
which is also symmetric and negative definite.
 The first equation implies $z = - b\,A^{-1}\,y$, which gives with the second relation $1 = b\,(a - \,^tx\,A^{-1}\,y)$ and thus with our assumptions 
\[ 
a -  \,^ty\,A^{-1}\,y > 0, \quad \quad
b = \frac{1}{a - \,^ty\,A^{-1}\,y}, \quad \quad
z = -  \frac{1}{a - \,^ty\,A^{-1}\,y}\,A^{-1}\,y.
\]
The third and fourth equations are then satisfied with the symmetric matrix
\[
B = A^{-1} +   \frac{1}{a - \,^ty\,A^{-1}\,y}\,
(A^{-1}\,y)\,^t(A^{-1}\,y).
\]
In terms of the positive definite matrix $W = - A^{-1}$ the
condition $^tu\,B\,u < 0$ for $u \neq 0$ translates into 
\[
(^tu\,W\,y)^2 <  (^ty\,W\,y)\,(^tu\,W\,u) +  a\,(^tu\,W\,u).
\]
The Cauchy-Schwarz inequality with $u = y$  implies that $a > 0$.
$\Box$

\vspace{.2cm}

Compare the matrix equation above with the relation 
$\bar{k}_{\alpha \beta}\,\bar{k}^{\beta \gamma} 
= \delta_{\alpha}\,^{\gamma}$. 
Since $S$ is space-like and $\bar{k}_{AB} = g_{AB}$  on $S_c$,  $\bar{k}_{AB}$ is negative definite. We saw above $\bar{k}^{00} > 0$ on $S_c$. Thus 
$\bar{k}^{AB}$ is negative definite if and only if 
$\bar{k}_{00} =
g_{00} + 2\,g_{3 (0}\,\phi_{,0)} +
g_{33}\,\phi_{,0}\,\phi_{,0} > 0$ on $S_c$, or
\begin{equation}
\label{k00>0-cond}
\bar{k}_{00} = g_{\mu\nu}\,P^{\mu}\,P^{\nu} > 0
\quad \mbox{on $S_c$ with} \quad 
P^{\mu} = \delta^{\mu}\,_0 + \phi_{,0}\,\delta^{\mu}\,_3.
\end{equation}
It holds $N_{\mu}\,P^{\mu} = 0$. Thus $P^{\mu}$ is tangential to $T_c$ on $S_c$, but  without further assumptions it need not be time-like.
By suitable choices of  the 
hypersurface ${\cal T}$ and the coordinates $z^{\mu}$, which were rather arbitrary so far, it can be arranged that
$P^{\mu}$ is in fact  time-like on, whence near $S_c$ with $\phi_{,0}$ satisfying (\ref{phi,0-range}).

\vspace{.2cm}

The Cauchy problem for the system 
(\ref{e-x-evol}), (\ref{short-phi-equ}), (\ref{phi,alpha-equ}),
(\ref{De0-gamma-x-a'-equ}), (\ref{De0-D-gamma-e-0-equ}),  
(\ref{De0-D-gamma-e-A-equ}) with initial data on open subsets of $S_c$ as discussed above is then well posed. The corresponding solutions define pieces of the prospective hypersurface
$T_c$. These local pieces can be patched together to obtain a part of the hypersurface $T_c$ diffeomorphic to $[0, x^{0'}_*[ \times S_c$, where $x^{0'}_* > 0$ and the unknown $x^{0'}$ takes values in $[0, x^{0'}_*[$. 

Because these solutions depend smoothly on the initial data and,  for sufficiently small $c_* > 0$, on $c \in [0, c_*[$, the hypersurfaces $T_c$ define a foliation that is smooth in the sense that the function $x^{3'}$ obtained by setting $x^{3'} = c$ on $T_c$ is smooth. Moreover, the coordinates $x^{\alpha'}$ and the vector fields $e_0$, $e_A$ obtained on the patches can be glued together to give smooth coordinates and vector fields (expressed in terms of 
$x^{\alpha'}$) that satisfy equations (\ref{e-x-evol})
with the given functions $F^A = F^A(x^{\alpha'}, x^{3'})$
on the domain covered by the foliation.
The mean extrinsic curvature of the $T_c$ is given there by
$\chi = \chi(x^{\alpha'}, x^{3'})$ (or $f = f(x^{\alpha'}, x^{3'})$).

\section{Concluding remarks}

It has been shown that the gauge based in section 2 on the choice of the function $x^3$ and the vector field $e_0$, both suitably  adapted to the given time-like boundary $T$, can be completely reconstructed, together with the hypersurface $T$, on the basis of the given gauge source functions $f$ and $F^A$. The construction imposes no conditions on the underlying space-time and, in particular, does not require the metric to satisfy any field equation.
The system of differential equations required for this turns out to be fairly complicated and only quasi-linear. Obtaining information about the life time of a gauge is a notoriously difficult problem. The characterization in terms of $f$ and $F^A$ appears particularly difficult.
A closer comparison with the way the gauge is discussed in section 2, which covers without complications a whole neighbourhood of the {\it given} hypersurface $T$, may give some insight into this.

Because the system considered in section 4 looks so difficult, it may be mentioned that  fixing the gauge in terms of $f$ and $F^A$
introduces no additional complication into the reduced system extracted in \cite{friedrich:nagy} from equations (\ref{torf}), (\ref{Gcurv}), (\ref{frvacbian}).
Part of the reason is that the curvature, which somewhat  unexpectedly enters the system derived in section 4, is already an unknown in the reduced system. In a similar way as 
equation (\ref{gauge-equ}) reduces to the relation 
$- g^{\nu \lambda}\,\Gamma_{\nu}\,^{\mu}\,_{\lambda}
= F^{\mu}$ if it is expressed in terms of the coordinates $x^{\mu}$ that solve the equation, the equations of section 4 reduce to simpler expressions if they are expressed in terms of their solution so that ${\cal T}$ coincides with the hypersurface $T_0$ and $x^{3'} = z^3$ whence $\phi = 0$. The possibility to declare a particular set of functions as gauge source functions and the usefulness of this choice obviously
depends on the chosen representation of the field equations.

}


\begin{thebibliography}{11}

\bibitem{bartnik:1988b}
R. Bartnik.
\newblock 
Remarks on cosmological space-times and constant mean curvature surfaces.
\newblock {\em Commun. Math. Phys.} 117 (1988) 615 - 624.

\bibitem{brill:1982}
D. Brill.
\newblock On spacetimes without maximal surfaces.
\newblock In: Proc. Third Marcel Grossman meeting, Ning, H. (ed.). Amsterdam: North-Holland 1982

\bibitem{fournodavlos-smulevici:2020}
G. Fournodavlos, J. Smulevici
\newblock The initial boundary value problem for the Einstein equations with totally geodesic time-like boundary.
\newblock arXiv:2006.01498 


\bibitem{friedrich:1hyp red}
H. Friedrich.
\newblock On the hyperbolicity of Einstein's and other gauge field 
equations.
\newblock {\it Comm. Math. Phys.} 100 (1985) 525--543.


\bibitem{friedrich:AdS}
H. Friedrich.
\newblock Einstein equations and conformal structure: existence of
anti-de Sitter-type space-times.
\newblock {\it J. Geom. Phys.}, 17 (1995) 125--184.

\bibitem{friedrich:ibvp-unique}
H. Friedrich.
\newblock Initial boundary value problems for Einstein's field equations and geometric uniqueness.
\newblock {\em Gen Relativ Gravit} 41 (2009) 1947 - 1966.






\bibitem{friedrich:nagy}
H. Friedrich, G. Nagy.
\newblock The initial boundary value problem for Einstein's vacuum 
field equations.
\newblock {\it Comm. Math. Phys.} 201 (1999) 619 - 655.

\bibitem{friedrichs:sym-hyp}
K. O. Friedrichs.
\newblock Symmetric hyperbolic linear differential equations.
\newblock {\it Comm. Pure and Appl. Math} 11 (1954) 
345 - 392.


\bibitem{holst:nagy:tsogtgerel:2008}
M. Holst, G. Nagy, and G. Tsogtgerel.
\newblock Far-from-constant mean curvature solutions of Einstein's constraint equations
 with positive Yamabe metrics.
\newblock {\it PRL} 100 (2008)  161101.

\bibitem{lindblom+al:2006}
L. Lindblom, M. Scheel, L. Kidder, R. Owen,
O. Rinne.
\newblock A new generalized harmonic evolution system.
\newblock {\em Class. Quantum Grav.} 23 (2006) S447 - S462.



\bibitem{maxwell:2009}
D. Maxwell.
\newblock A class of solutions of the vacuum Einstein constraint equations with freely specified mean curvature.
\newblock {\em Math. Res. Lett.} 16(4)  (2009) 627 - 645.



\bibitem{pretorius:2005a}
F. Pretorius. 
\newblock Evolution of binary black hole spacetimes. 
\newblock {\em Phys. Rev. Lett.} 95 (2005) 21101. 

\bibitem{pretorius:2005b}
F. Pretorius. 
\newblock Numerical relativity using a generalized harmonic decomposition. 
\newblock {\em Class. Quantum Grav.} 22 ( 2005) 425 - 452.













\bibitem{ringstroem:2008}
H. Ringstr\"om.
\newblock Future stability of the Einstein-non-linear scalar field system.
\newblock {\em Invent. math. }173  (2008) 123 - 208.



\end{thebibliography}
\end{document}